# Gradients in solid surface tension drive Marangoni-like motions in cell aggregates

Vikrant Yadav,[1,2,*] Md. Sulaiman Yousafzai,[1,2,5,*] Sorosh Amiri,[2,3] Robert W. Style,[4] Eric R. Dufresne,[4] and Michael Murrell[1,2,4,5,†]

[1]*Department of Biomedical Engineering, Yale University, 55 Prospect Street, New Haven, Connecticut 06511, USA*
[2]*Systems Biology Institute, Yale University, 850 West Campus Drive, West Haven, Connecticut 06516, USA*
[3]*Department of Mechanical Engineering and Material Science, Yale University, 10 Hillhouse Avenue, New Haven, Connecticut 06511, USA*
[4]*Department of Materials, ETH Zurich, Zurich 8093, Switzerland*
[5]*Department of Physics, Yale University, 217 Prospect Street, New Haven, Connecticut 06511, USA*

The surface tension of living cells and tissues originates from the generation of nonequilibrium active stresses within the cell cytoskeleton. Here, using laser ablation, we generate gradients in the surface tension of cellular aggregates as models of simple tissues. These gradients of active surface stress drive large-scale and rapid toroidal motion. Subsequently, the motions spontaneously reverse as stresses reaccumulate and cells return to their original positions. Both forward and reverse motions resemble Marangoni flows in viscous fluids. However, the motions are faster than the timescales of viscoelastic relaxation and the surface tension gradient is proportional to mechanical strain at the surface. Further, due to active stress, both the surface tension gradient and surface strain are dependent upon the volume of the aggregate. These results indicate that surface tension can induce rapid and highly correlated elastic deformations in the maintenance of tissue shape and configuration.

Whenever there is a gradient in surface tension along an interface, there is an associated shear stress. For the interfaces of fluids, this results in a flow near the interface that converges on regions of high surface tension, known as the Marangoni effect [1]. One classic example is "tears of wine" [2]. Marangoni stresses are also well known to cause convection in thin, heated films, driven by local temperature gradients along the film surface, and arise whenever there are surfactant or solute gradients [3]. While Marangoni effects have been studied extensively in simple fluids, and to some extent in viscoelastic fluids [4–6], their impact on solids appears to be unexplored. In part, this is because surface tension driven deformations of solids are *typically* limited to length scales smaller than the elastocapillary length $l = \gamma/E$ (where $\gamma$ is the surface tension and $E$ is Young's modulus) [7]. However, in living systems, the magnitude of stresses caused by cell and tissue surface tensions may be comparable to their bulk elasticity on cellular length scales [8,9]. Thus, capillary effects may play a large role in their mechanical behaviors.

Tissue surface tension originates in the nonequilibrium activity of mechanochemical enzymes within the cell cytoskeleton that convert chemical energy into "active" stress [10–12]. At the cellular level, these stresses contribute to cortical and membrane tensions that determine cell shape and mediate their response to mechanical perturbation [13–17]. At large scales, cortical tensions are coupled together via cell-cell adhesions and lead to tissue surface tensions [16,18–20]. In the bulk of the tissue, stresses may lead to a heterogenous pressure profile and reorganize the distribution of cells [21–25]. Tissue surface tension has been considered liquidlike on timescales of cell migration, in which the turnover of cell-cell adhesion relaxes internal stress [26]. On shorter timescales, the surface tension may be solidlike [27]. Likewise, the effects of gradients in solid surface tension may impact cell motion.

---

*These authors contributed equally to this work.
†michael.murrell@yale.edu

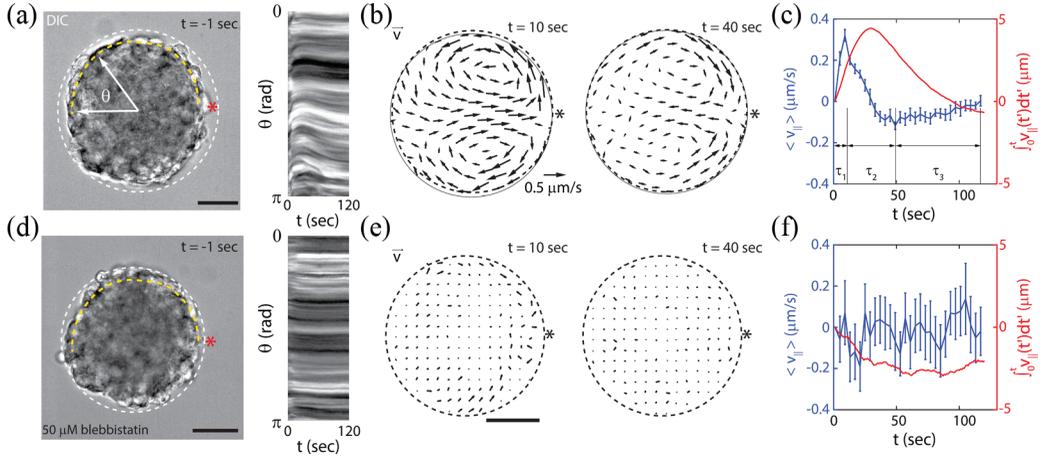

FIG. 1. Cell ablation at the aggregate surface induces rapid, internal toroidal motions. (a) Left: DIC image of a cell aggregate immediately prior to ablation. The location of the ablation is at the red asterisk. Time is time after ablation ($t = 0$ sec). Right: Kymographs depicting material displacement around the cell surface. The yellow dashed line indicates the path over which kymographs are calculated. The white dotted line indicates the region over which PIV is applied. (b) PIV of cellular motion ($\vec{v}$) analyzed from DIC images, taken at two different time points. The time between frames is 1 sec. (c) Mean speed and net material transport over time. The timescales of relaxation and reversal are denoted by $\tau_1$, $\tau_2$, and $\tau_3$. (d) Left: Image of a cell aggregate. Right: Corresponding kymographs with 50 $\mu$M Blebbistatin. (e) PIV of cellular displacements with 50 $\mu$M Blebbistatin. (f) Mean speed and net material transport for Blebbistatin treated aggregate. Scale bars are 30 $\mu$m.

In this Letter, we show that surface tension gradients can drive rapid and large-scale collective motions within cellular aggregates, as models of simple tissues. We use laser ablation to quickly and locally reduce aggregate surface tension, creating shear stresses at the surface. This drives large-scale cell movements which reverse as surface tension reaccumulates. These motions include multiple timescales, which are compared to the viscoelastic relaxation timescale to determine the extent to which they can be considered elastic or viscous. To further make this determination, we measure the strain in the cells at the aggregate surface and compare it to the estimated surface tension gradient to define an effective surface modulus. Finally, we inhibit the conversion of chemical energy into active stress by myosin II molecular motors to understand the role of nonequilibrium processes in the aggregate mechanical response.

Cell aggregates are composed of S180 murine sarcoma cells, formed from suspension spinning, and are 50–150 $\mu$m in diameter (see Supplemental Material, Fig. 1) [28,29]. E-cadherin interactions induce cell-cell adhesion, leading to round aggregates with a smooth surface [Fig. 1(a)]. To generate gradients in surface tension, we use a high-energy laser ($\lambda = 337$ nm, 40 mW) to ablate cells at a point at the surface of the nonadhered cell aggregate (Fig. 1(a) and Supplemental Material, Video 1 [29]). As the laser is focused at a 0.6 $\mu$m diameter area (2X objective, NA 0.8), the number of cells removed during ablation is few and consistent across different experiments. We quantified the resultant time-resolved displacements using particle image velocimetry (PIV) (Supplemental

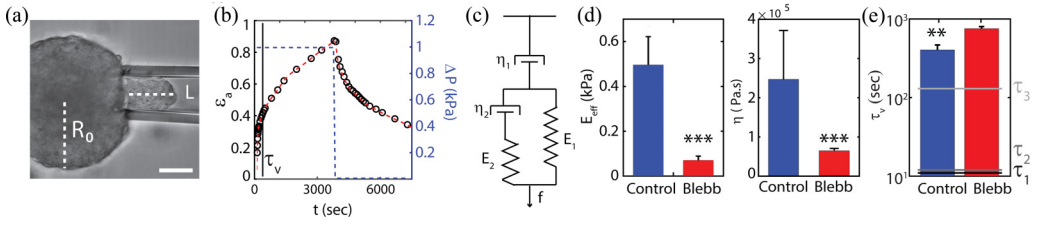

FIG. 2. The mechanical relaxation is slow compared to the timescales of motion. (a) Bright-field image of an aggregate in the micropipette. (b) A step in pressure ($\Delta P$) is applied to a cell aggregate of radius $R_0$ that brings the aggregate into the pipette, a length $L$ and induces a strain, $\varepsilon_a$, as a function of time. The solid black line represents $\tau_v$. (c) Modified Maxwell model applied to the evolution of $\varepsilon_a$ with time. In the model, $E_1$ and $E_2$ are elastic modulus, and $\eta_1$ and $\eta_2$ are dashpots representing dissipative elements. $f$ represents the stress applied to the system, in analogy to the applied pressure by the micropipette. (d) The effective elastic modulus $E_{\text{eff}}$ and viscosity ($\eta$) obtained from fitting experimental data. (e) Viscoelastic relaxation time $\tau_v$ of aggregates for an applied pressure of 1 kPa, in comparison with motion timescales $\tau_1$, $\tau_2$, and $\tau_3$ as defined in Fig. 1(c) ($n = 10$). Scale bar is 30 $\mu$m. *** $p < 0.001$, ** $p < 0.01$.

Material, Note 1 [29]). Using PIV measurements, we experimentally measure the flow profiles, calculate the shear strains at the surface, and use the strains to calculate gradients in surface tension.

Immediately upon ablation, the surface of the aggregate retracts from the ablation site [Fig. 1(b)]. The rearward surface motions are concomitant with immediate forward motions of the core (center). In combination, these motions yield toroidal patterns that resemble Marangoni flows in liquid droplets (Supplemental Material, Fig. 2 [29]) [30,31]. There are multiple timescales within the observed motions. The timescale of the initial relaxation is rapid ($\tau_1 = 10$ sec). Subsequently, the surface and bulk motions reverse. Thus, the toroid reverses its orientation over two timescales: first quickly ($\tau_2 = 40$ sec) and then slowly as the aggregate approaches its original configuration ($\tau_3 = 70$ sec) [Fig. 1(c)]. To estimate the extent of reversibility in motion and how much the aggregate reestablishes its original configuration, we calculate the net material displacement through the center of the aggregate. The net material transport can be approximated as $\int_0^t v(t')dt'$, where $v$ is the velocity of material motion at the center of the aggregate. The net material displacement increases with forward motion and decreases with reversal, finally returning to zero, indicating the motion is reversible [Fig. 1(c)]. Both forward and reverse motions are dependent upon active stress, as aggregates treated with 50 $\mu$M Blebbistatin, which inhibits myosin II ATPase activity, show no coherent motion upon ablation (Figs. 1(d)–1(f), and Supplemental Material, Video 2 [29]).

To compare the observed timescales of motion and reversal with the viscoelastic timescale of the aggregate, we quantify the mechanical properties of the aggregate with micropipette creep and stress relaxation tests (Figs. 2(a) and 2(b), and Supplemental Material, Video 3, Note 2, and Fig. 3 [29]). Briefly, a micropipette of 30 $\mu$m radius is used to apply a suction pressure on the aggregate. This draws a length $L$ of aggregate into the pipette, which increases over time until the pressure is released (here, after approximately 60 min), at which point $L$ decreases. The changes in $L$ with time reflect the viscoelasticity of the aggregate. Following previous work [32], the aggregate can be well described by the modified Maxwell model (Fig. 2(c), and Supplemental Material, Figs. 4 and 5 [29]),

$$\varepsilon_a = \frac{f}{\eta_1}t + \frac{f}{E_1}\left(1 - \frac{E_2}{E_1 + E_2}e^{-\frac{E_1 E_2}{\eta_2(E_1+E_2)}t}\right). \tag{1}$$

Here, $f$ is the applied micropipette stress, the strain $\varepsilon_a$ is defined as $L_t/R_0$, where $R_0$ is the undeformed radius of the aggregate, and $E_1$, $E_2$, $\eta_1$, and $\eta_2$ are viscoelastic constants of the modified Maxwell model. We define a viscoelastic timescale, $\tau_v = \eta_2/E_{\text{eff}}$, where $E_{\text{eff}}$ is given by $E_1 E_2/(E_1 + E_2)$. Thus, at short times ($t \ll \tau_v$), the aggregate behaves as an elastic solid. By

VIKRANT YADAV *et al.*

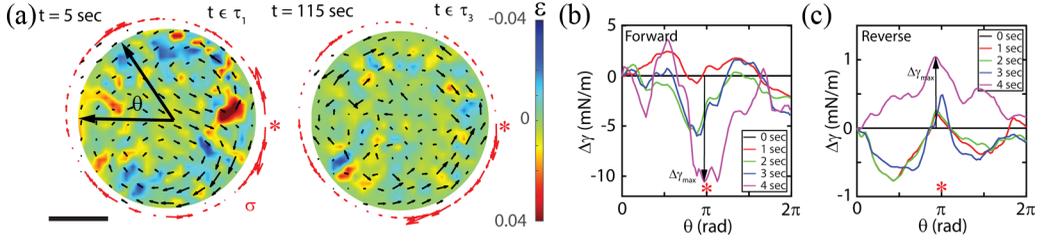

FIG. 3. Motions are driven by solidlike surface stresses. (a) Cell displacement $\vec{u}$ (black vectors), areal strain field ($\varepsilon = \nabla \cdot \vec{u}$) due to displacement (color map), and surface stresses (red vectors) associated with deformation at $t = 5$ sec (during $\tau_1$) and $t = 115$ sec (during $\tau_3$). (b) Change in surface tension in forward and (c) reverse motion, as a function of angular position $\theta$, as measured from constitutive models. Asterisks indicate the region of ablation. The black arrow in (b) and (c) denotes the maximum change in surface tension at a given instant. Scale bar is 30 $\mu$m.

contrast, at long times ($t \gg \tau_v$), the aggregate flows like a liquid with viscosity $\eta_1$ (see Supplemental Material, Figs. 4 and 5, Note 3 [29]). This is consistent with previous studies that suggest that cell aggregates behave as solids at short timescales and as fluids at long timescales [33–38].

We estimate the values of the associated material parameters and timescales by fitting aspiration and retraction curves to Eq. (1). We find that $E_1 = 700$ Pa, $E_2 = 1.7$ kPa, $\eta_1 = 2.5 \times 10^5$ Pa sec, and $\eta_2 = 2 \times 10^5$ Pa sec. Finally, using these fitted values, we get $\tau_v = 402$ sec $+/-$ 65 sec. Further, like the timescales of motion, we find that the magnitude of all the parameters are dependent on myosin activity, as all are reduced upon treatment with 50 $\mu$M Blebbistatin [39] [Fig. 2(d)]. Thus in the presence of active stress, we note that the timescales of motion as described in Fig. 1 are all shorter than the viscoelastic timescale as measured by micropipette [Fig. 2(e)]. We therefore conclude that all motions within the aggregate bulk are predominantly elastic.

Given these bulk rheological parameters, we can infer the surface properties that drive motion from the analysis of the deformations. For a single aggregate, both forward and reverse strains are measured by PIV [Fig. 3(a)]. For this, we calculate the displacements for $\Delta t = 4$ sec in the forward direction ($t = 0$ sec to $t = 4$ sec, during $\tau_1$), and for $\Delta t = 4$ sec in the reverse direction ($t = 120$ sec to $t = 116$ sec, during $\tau_3$). This time interval is the longest time over which the vector field can be estimated using PIV for forward motion, and is used for the reverse motion for comparison. As $\Delta t \ll \tau_v$ [Figs. 1(c) and 2(e)], we assume a linear elastic constitutive relationship and multiply the strain by the modulus as measured by micropipette to estimate the surface stress. We use the shear stress measurements to calculate changes in surface tension along the aggregate periphery [40] (see Supplemental Material, Note 4 and Fig. 6 [29]). In particular, shear stresses at the surface, $\sigma$, must be balanced by a corresponding gradient in the surface tension, which is given by the following relation:

$$\sigma = G\varepsilon_s = d\gamma/ds = \frac{1}{R_0}\frac{d\gamma}{d\theta}, \qquad (2)$$

where $\varepsilon_s$ is the shear strain, $G$ is the shear modulus, $s$ is the arc length along the aggregate surface, $\theta$ is the corresponding angular position, and $R_0$ is the radius of the aggregate. $R_0$, $\theta$, and $s$ are related by the constraint $s = R_0\theta$. The calculation of shear strain $\varepsilon_s$ from the PIV data is outlined in the Supplemental Material, Note 4 [29], and shear modulus $G$ is approximated as $E_1$. We integrate (2) with respect to $\theta$ to get the changes in surface tension, $\Delta\gamma$, relative to a fixed starting point. Equation (2) assumes force balance at the aggregate surface and does not require calculation of bulk stresses. The grid resolution used in PIV is 3.2 $\mu$m, which is comparable to the size of a cell. Therefore, the calculated surface tensions pertain to a single cell layer at the surface of the aggregate. Relative to the initial time (forward motion), plots of $\Delta\gamma$ [Fig. 3(b)] show a localized drop in surface tension near the ablation point of approximately 10 mN/m, which drives toroidal motion away from this point. Relative to the final time point (reverse motion) [Fig. 3(c)], over the same time interval, we see a localized increase of approximately 1 mN/m, which pulls cells back in the opposite direction.

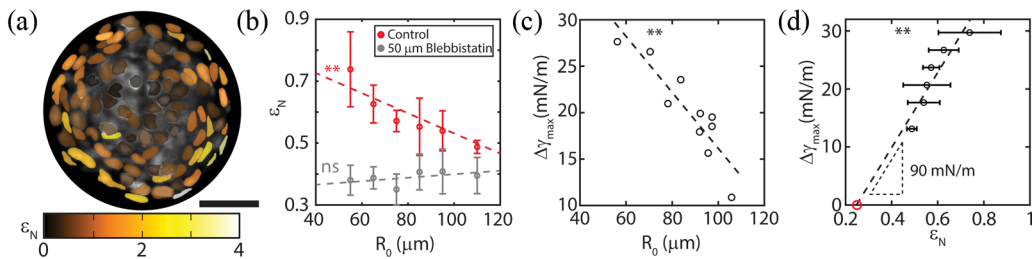

FIG. 4. Surface tensions are size dependent and solidlike. (a) Distribution of nuclear shapes inside an aggregate color coded by their strain defined as $\varepsilon_N = AR - 1$. Scale bar is 30 $\mu$m. (b) Average nuclear strain ($n = 36$) varies with the size of the aggregate $R_0$. (c) Surface tension gradient $\Delta\gamma$ ($n = 10$) varies with the size of the aggregate $R_0$. (d) Surface tension gradient vs nuclear strain. The red dot indicates nuclear strain in the presence of 50 $\mu$M Blebbistatin. The slope of the curve represents the apparent surface modulus. ** $p < 0.01$ ns is not significant.

To understand the origin of the surface tension, we looked for changes to the cell morphology with aggregate size (volume). Specifically, we imaged the deformation of cell nuclei with a nuclei-localizing green fluorescent protein [GFP, Fig. 4(a)]. We observe that the nuclei closer to the edge of the aggregate are significantly more elongated, whereas the ones closer to the center are more rounded. The observed radial anisotropy is consistent with previous results [21,41]. We quantify the extent that the nuclei are elongated compared to circular in shape, and define a strain $\varepsilon_N = AR - 1$, where AR is defined as the ratio of the major and minor axes of an ellipse fitted onto the nuclei.

For aggregates of different sizes, we observe that the strain decreases monotonically with the size of the aggregate [Fig. 4(b)]. Similarly, we quantify the surface tension difference $\Delta\gamma_{\max}$ during the initial retraction phase, $\tau_1$. This represents the largest change in surface tension during the motion of the aggregate after ablation. Indeed, we observe that cellular aggregates demonstrate the presence of a size-dependent $\Delta\gamma_{\max}$ [Fig. 4(c)]. The difference is largest for small aggregates and decreases identically as strain for large aggregates.

We attribute the decrease in nuclear strain ($\varepsilon_N$) as well as the decrease in stress to a decrease in the density of phosphorylated myosin on the aggregate surface (Supplemental Material, Fig. 7 [29]). Further, we find that like the phosphorylation of myosin, the strain and the stress decrease monotonically with the size of the aggregate. Therefore, we suggest that there are strong active stresses at the surface, although the existence of bulk stresses cannot be excluded. However, suppression of motor activity by Blebbistatin suppresses all size-dependent effects (Fig. 4(b), and Supplemental Material, Fig. 7 [29]).

The covariation of surface tension and nuclear strain with aggregate dimensions can be rationalized through a strain-dependent surface tension. Plotting $\Delta\gamma$ versus the surface strain, $\varepsilon_N$, which were independently plotted functions of aggregate size $R_0$, we find a linear relationship between surface tension and nuclear strain [Fig. 4(d)]. This slope of this surface constitutive relationship defines an apparent surface modulus, $E_s = 90$ mN/m approximately. This value is comparable to other soft solids, such as silicone gels [42–44] and cell membranes [45].

Intriguingly, the stress-strain curve extrapolates with a straight line to a nonzero $\Delta\gamma_{\max}$ at zero strain. We attribute this residual strain in the aggregate to the non-myosin-based components of surface tension (termed "passive" for simplicity) which are unperturbed by laser ablation. As Blebbistatin-treated aggregates do not flow after ablation, it is not possible to estimate surface stresses using the method described earlier. Therefore, passive contributions are not included in our estimate of $\Delta\gamma$. However, the aggregates retain a roughly spherical shape, indicative of passive surface stresses, likely emerging from cell-cell cohesion [46]. To provide additional evidence of the presence and magnitude of passive contributions to the surface tension, we measure the surface tension at long times by micropipette ($t \gg \tau_v$). In this case, we find that the surface tension for aggregates with active myosin is approximately 12.5 mN/m, in agreement with the surface tension gradient measured after ablation (Supplemental Material, Fig. 8 [29]). However, upon myosin inhibition, we find a surface tension of approximately 4 mN/m. Thus, we suggest that passive contributions to the surface tension are approximately 30%.

The toroidal pattern of motion observed in our laser ablation experiments is a hallmark of Marangoni flows. However, there are several principal distinctions from traditional Marangoni flows.

First, the observed toroidal displacements are distinct from traditional Marangoni flows in that the timescales of motion are short compared to the characteristic viscoelastic timescale, suggesting they are elastic in nature. Further, the surface tension gradient that generates the displacement is strain dependent, a hallmark of elastic solids. Thus, we conclude that the surface tension of aggregates on timescales on which toroidal motions are induced contains solidlike characteristics. While Marangoni flows are typically associated with surface tension-driven motion of fluids, the nature of the force balance remains the same for a linear elastic material. Indeed, one can be transformed to another by substituting the shear modulus with shear viscosity and displacements with velocities [47]. This equivalence between linear elasticity and the Stokes flow within the force balance indicates that surface tension gradients can equivalently drive Marangoni flows in fluids, or elastic Marangoni-like deformations in solids (Supplemental Material, Note 5 [29]).

Second, after the flows that relax the initial applied surface tension, as observed in passive droplets, the flows spontaneously reverse direction—a behavior unseen in passive systems. In doing so, aggregates in suspension return to their original shape and configuration. Elastic stresses that were relaxed are quickly reaccumulated using myosin-based active stresses. Thus, the aggregate may maintain "surface-tensional homeostasis" [48–50], a principle by which the system desires to restore a set level of tension after perturbation, analogous to tensional homeostasis previously reported in cells and tissues [48,51–55].

Third, the active surface tension gradient and surface strain are dependent upon the size (volume) of the aggregate. The measured volume dependence of the aggregate surface tension is consistent with the increased phosphorylation of myosin molecular motors at the surface (Supplemental Material, Fig. 7 [29]). In the absence of actomyosin activity, this size dependence is lost, consistent with the constant surface tension observed in passive systems, such as liquid droplets. Similar size dependence is seen in other systems, for example, in the actomyosin-driven compaction of mesenchymal stem cell volume [56,57] or the actomyosin assembly for purse string in epithelial sheets [58]. Here, the volume dependence yields a range of stresses and strains, and thus allows us to relate the two directly and define an active, effective surface modulus.

Finally, Marangoni flows observed in passive fluids are driven exclusively by surface tension gradients. Here, activity contributes to surface stresses, but may also contribute to bulk stresses and the accumulation of pressure within the aggregate. As a result, the elevation of surface tension by internal pressure (e.g., hoop stress) cannot be excluded. Therefore, to be inclusive of this possibility, we term the observed phenomenon presented here "Marangoni-like" motion.

In summary, cell aggregates, as models of simple tissue, have active and solidlike material properties, which can drive rapid and correlated internal cell motions.

We acknowledge funding through Grant No. ARO MURI W911NF-14-1-0403 to M.M. and V.Y., as well as Grant No. NIH RO1 GM126256 to M.M., and Grant No. NIH U54 CA209992 to M.M. and M.S.Y. M.M. also acknowledge support from Human Frontiers Science Program (HFSP) Grant No. RGY0073/2018. Any opinions, findings, and conclusions or recommendations expressed in this material are those of the authors(s) and do not necessarily reflect the views of the NIH or HFSP. We would like to thank Professor Erdem Karatekin for help with the micropipette measurements. We would also like to thank Professor Kate Jensen, Dr. Karine Guevorkian, and Professor Francoise Brochard-Wyart for constructive discussion.